# The second law, maximum entropy production and Liouville's theorem

**Roderick C Dewar[1] and Amos Maritan[2]**

[1]Research School of Biology, The Australian National University, Canberra ACT 0200, Australia
[2]Department of Physics G. Galilei, University of Padova, INFN, Via Marzolo 8, 35131 Padova, Italy
E-mail: roderick.dewar@anu.edu.au and amos.maritan@pd.infn.it

**Abstract.** In 1965 Jaynes provided an intuitively simple proof of the 2$^{nd}$ law of thermodynamics as a general requirement for any macroscopic transition to be experimentally reproducible. His proof was based on Boltzmann's formula $S = k\ln W$ and the dynamical invariance of the phase volume $W$ for isolated systems (Liouville's theorem). Here Jaynes' proof is extended to show that Liouville's theorem also implies maximum entropy production (MaxEP) for the stationary states of open, non-equilibrium systems. According to this proof, MaxEP stationary states are selected because they can exist within a greater number of environments than any other stationary states. Liouville's theorem applied to isolated systems also gives an intuitive derivation of the fluctuation theorem in a form consistent with an earlier conjecture by Jaynes on the probability of violations of the 2$^{nd}$ law. The present proof of MaxEP, while largely heuristic, suggests an approach to establishing a more fundamental basis for MaxEP using Jaynes' maximum entropy formulation of statistical mechanics.

**Keywords:** stationary states, stochastic processes (theory)

1. Introduction

The conjecture of Maximum Entropy Production (MaxEP) for non-equilibrium stationary states has shown some predictive successes in studies of, for example, planetary climates [1, 2], turbulent fluids [3, 4], crystal growth morphology [5, 6] and biological design [7, 8]. Potentially, MaxEP is a selection principle of great generality [9]. However, the theoretical basis for the empirical success of MaxEP remains unclear, and this has limited its wider application. An attempt to understand MaxEP was made by one of us [10, 11], by applying Jaynes' maximum entropy (MaxEnt) formulation of statistical mechanics [12, 13] to the probability distribution of microscopic paths in phase space.

MaxEnt is a generic algorithm for constructing a Bayesian probability distribution $p_i$ from available information, by maximising the Gibbs-Shannon entropy $H = -\sum_i p_i \ln p_i$ with respect to $p_i$ subject to various moment constraints on $p_i$ which represent the available information. The origins of MaxEnt go back to Gibbs' formulation of equilibrium statistical mechanics [14, 15], and an information-theoretical rationale for MaxEnt was developed subsequently by Jaynes and others [12, 13, 15-17]. While the specific arguments used to derive MaxEP from MaxEnt in [10, 11] have been rightly criticised on technical grounds [18-21], the basic MaxEnt approach remains of interest, in part because it leads to well-established results such as the fluctuation theorem [10, 11, 22, 23]. Moreover, because of its generality, which extends from equilibrium to non-equilibrium statistical mechanics and beyond [17], MaxEnt is still thought to offer a promising basis for understanding the empirical success of MaxEP across a wide range of disciplines, see e.g. [21, 24, 25].

In this paper we wish to present an alternative heuristic explanation of MaxEP based on the dynamics of phase volumes for isolated systems, from which it is possible to see along what lines a more fundamental derivation of MaxEP from MaxEnt might proceed. This heuristic derivation of MaxEP is a simple extension of an argument used by Jaynes to prove the $2^{nd}$ law of thermodynamics [26, 27].

MaxEP can be viewed as a codicil to the $2^{nd}$ law. The $2^{nd}$ law governs the overall direction of macroscopic processes but does not say anything about their rates; MaxEP purports to predict their rates. Jaynes [26, 27] provided an intuitively simple proof of the $2^{nd}$ law as a general requirement for any macroscopic transition to be experimentally reproducible. His proof was based on Boltzmann's formula $S = k\ln W$ and the dynamical invariance of the phase volume $W$ for isolated systems (Liouville's theorem). Here we extend Jaynes' proof of the $2^{nd}$ law to show that Liouville's theorem also implies MaxEP for the stationary states of open, non-equilibrium systems.

A wider motivation, alluded to above, for establishing the present heuristic explanation of MaxEP is that it might suggest how to base a more fundamental understanding of both the $2^{nd}$ law and MaxEP on MaxEnt, rather than on phase volumes. This possibility arises from the fact that the Gibbs-Shannon entropy $H$ has an interpretation in terms of phase volumes which generalises Boltzmann's formula [26]. This link between $H$ and phase volume, together with the phase volume explanation of MaxEP presented here, then suggests how MaxEP might be derived more generally from MaxEnt.

The structure of the paper is as follows. Section 2 presents Jaynes' proof of the $2^{nd}$ law in terms of phase volumes, within an extended picture of microscopic paths and their conjugate reverse paths (figure 1). This phase volume picture sets the scene for the subsequent derivations of the fluctuation theorem in section 3 and MaxEP in section 4. Section 5 concludes with an outline of the phase volume interpretation of MaxEnt (generalised Boltzmann formula), and the implications of the results of sections 3 and 4 for a more fundamental understanding of MaxEP based on MaxEnt.

## 2. Jaynes' proof of the second law of thermodynamics from Liouville's theorem

Consider an isolated system, which we will subsequently resolve into an open sub-system plus its environment (section 4). The phase space of the isolated system is the set of its possible microscopic states. For example, a classical Hamiltonian system consisting of $n$ particles has a $6n$-dimensional phase space in which the microstates are specified by the positions and momenta of each particle, $\{q_i, p_i, i = 1 .. n\}$. For every microstate $\{q_i, p_i\}$ there exists a mirror microstate $\{q_i, -p_i\}$ having the same particle positions but the reverse particle velocities.

A macrostate is a less detailed description of the system. When considering non-equilibrium states, the macrostate description includes a specification of any macroscopic fluxes that are present. A given flux-specific macrostate can be realized by many different microstates. Figure 1 depicts three such macrostates: A, A' and B. The phase volume of macrostate A, denoted by $W(A)$, is the number of microstates that realize macrostate A; the microstates in region A are thus macroscopically equivalent. To each macrostate A there corresponds a mirror macrostate RA related by velocity reversal (denoted by the operator R, see figure 1). Clearly, $W(RA) = W(A)$. This mirror relationship is not confined to classical

Hamiltonian systems, but applies to all systems governed by time-reversible microscopic equations of motion. Similar statements hold for A′ and B.

The left part of figure 1 illustrates the situation where the regions A∪RA and B∪RB are dynamically linked. The path Γ represents a possible microscopic path (i.e. a solution of the microscopic equations of motion of the isolated system) during time interval (0, τ). Let Γ(*t*) denote the microstate of the (isolated) system at time *t* when it follows path Γ. Regions (macrostates) A and B are the set of microstates that are macroscopically equivalent, respectively, to Γ(0) and Γ(τ), i.e. Γ realizes the macroscopic transition A → B.

Γ starts from microstate Γ(0) ∈ A and ends at microstate Γ(τ) ∈ TA (where T denotes the time evolution operator over interval τ); Liouville's theorem (conservation of phase volumes by the microscopic equations of motion) implies $W(TA) = W(A)$ as depicted in figure 1. Velocity reversal of Γ(τ) defines microstate Γ*(0) ∈ RTA; starting from Γ*(0) the system follows path Γ* to microstate Γ*(τ) ∈ RA which is just the velocity reversal of Γ(0). Γ* defines the path conjugate to Γ; Γ* realizes the macroscopic transition RB → RA. The right part of figure 1 shows analogous transitions between regions B∪RB and A′∪RA′.

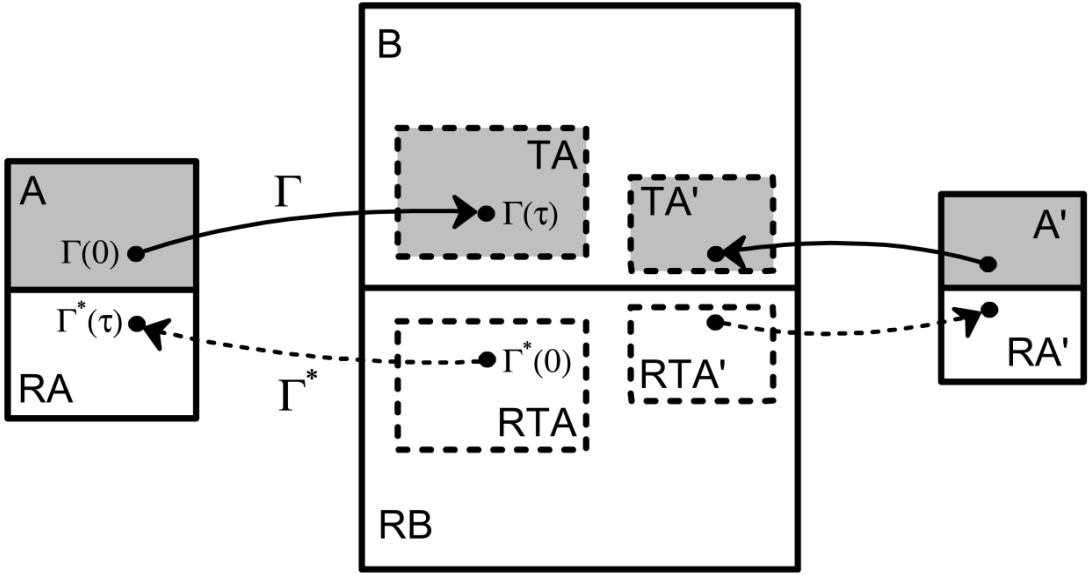

**Figure 1.** Left: Schematic representation of phase volume relationships associated with dynamic transitions between phase space regions A∪RA and B∪RB of an isolated system. R denotes the operation of velocity reversal. T denotes the time evolution operator over time interval τ. All microstates within region A are macroscopically equivalent, thus defining the flux-specific macrostate A (similarly for RA, B and RB). Γ is a microscopic path that takes the system from Γ(0) ∈ A to Γ(τ) ∈ TA during interval τ. Γ* is the microscopic path conjugate to Γ that takes the system from Γ*(0) ∈ RTA to Γ*(τ) ∈ RA. Liouville's theorem implies that A and TA have equal phase volumes. For the macroscopic transition A → B to be experimentally reproducible requires TA ⊆ B (which is the case illustrated here), implying $W(A) \leq W(B)$, i.e. the 2$^{nd}$ law. Right: analogous transitions between regions B∪RB and A′∪RA′.

Figure 1 illustrates the situation where the macroscopic transitions A → B and A′ → B are *experimentally reproducible* over time interval τ [26, 27]. That is, every time we set up the system in macrostate A or A′ at *t* = 0, the system ends up in macrostate B at *t* = τ, the reason being that TA and TA′ lie entirely within B (figure 1). Thus, although we cannot control by macroscopic means which of the $W(A)$ microstates realizes A at $t = 0$, the condition TA $\subseteq$ B ensures that the system will always be found in B at $t = \tau$ (and similarly for A′ → B).

The condition TA $\subseteq$ B for experimental reproducibility of the macroscopic transition A → B implies $W(TA) \leq W(B)$. But $W(TA) = W(A)$ (Liouville's theorem), and so experimental reproducibility of A → B implies $W(A) \leq W(B)$ or $S(A) \leq S(B)$, where $S(M) = k\ln W(M)$ is the Boltzmann entropy of macrostate $M$. Similarly, experimental reproducibility of A′ → B requires $S(A′) \leq S(B)$. If A → B and A′ → B are experimentally reproducible, then the reverse macroscopic transitions RB → RA and RB → RA′, for which $S(RB) > S(RA)$ and $S(RB) > S(RA′)$, are not experimentally reproducible because they require additional microscopic control to confine the initial microstate to lie within the respective subregions RTA and RTA′ of RB. Thus Liouville's theorem implies that the 2$^{nd}$ law is a general requirement for any macroscopic transition to be experimentally reproducible [26, 27].

## 3. Entropy production of microscopic paths and the fluctuation theorem

While the Boltzmann entropy $S(M) = k\ln W(M)$ is defined as a function of macrostates $M$, it can also be defined for any microstate $x$ as $S_x = k\ln W_x$ where $W_x$ is the volume of the region of phase space whose microstates are macroscopically equivalent to $x$. This leads to a natural definition of the mean rate of entropy production associated with a microscopic path $\Gamma$ over time interval $(0, \tau)$, as

$$\sigma_\Gamma \equiv \frac{k}{\tau} \ln \frac{W_{\Gamma(\tau)}}{W_{\Gamma(0)}} \tag{1}$$

where, as in section 2, $\Gamma(t)$ denotes the microstate of the (isolated) system at time $t$ when it evolves on path $\Gamma$. For example, the path $\Gamma$ shown in figure 1 has entropy production rate $\sigma_\Gamma = (k/\tau)\ln[W(B)/W(A)]$. For the conjugate path $\Gamma^*$ we have $\sigma_{\Gamma^*} = -\sigma_\Gamma$ since $W_{\Gamma^*(0)} = W(RB) = W(B)$ and $W_{\Gamma^*(\tau)} = W(RA) = W(A)$. For a path $\Gamma$ to realize an experimentally reproducible macroscopic transition requires $\sigma_\Gamma > 0$.

The phase volume relationships depicted in figure 1 give an intuitive explanation for why, if the macroscopic transition A → B is experimentally reproducible, the reverse transition RB → RA, though possible, is improbable: by macroscopic means we are only able to place the system in some uncontrolled point in macrostate RB, but the reverse transition RB → RA requires the stronger condition that the initial microstate lies in the subregion RTA of RB. Jaynes [28] conjectured that the probability of the reverse transition is 'something like' $W(A)/W(B)$, although he did not consider velocity reversal explicitly. This result is intuitively obvious from figure 1, however, because $W(A)/W(B) = W(RTA)/W(RB)$ is just the probability that the isolated system is in subregion RTA given that it is in macrostate RB (we have assumed that all microstates realizing a given macrostate are equally probable).

We can also use figure 1 to re-express Jaynes' result as a relationship between the ratio $p_{\Gamma^*}/p_\Gamma$ of the probabilities of conjugate paths $\Gamma^*$ and $\Gamma$ and the corresponding entropy production rate $\sigma_\Gamma$. Since the microscopic dynamics is deterministic, the probability $p_\Gamma$ is just the probability that the initial microstate at $t = 0$ is $\Gamma(0)$. Therefore, given that the initial macrostate is A, we have

$$p_\Gamma = \frac{1}{W_{\Gamma(0)}} = \frac{1}{W(A)} \qquad (2)$$

(again we have assumed that all microstates realizing A are equally probable). Similarly for the reverse transition, given that the initial macrostate is RB, we have

$$p_{\Gamma^*} = \frac{1}{W_{\Gamma^*(0)}} = \frac{1}{W(RB)} \qquad (3)$$

From equations (1)-(3), $W(RB) = W(B) = W_{\Gamma(\tau)}$ and $W(A) = W_{\Gamma(0)}$ we then have

$$\frac{p_{\Gamma^*}}{p_\Gamma} = \frac{W(A)}{W(B)} = e^{\frac{-\tau \sigma_\Gamma}{k}} \qquad (4)$$

which is just Jaynes' conjecture [28] re-expressed at the level of microscopic paths. From equation (4), the p.d.f. for the entropy production rate, defined by

$$p(\sigma) = \sum_\Gamma p_\Gamma \delta_{\sigma_\Gamma, \sigma} \qquad (5)$$

then satisfies the fluctuation theorem [22, 23]

$$\frac{p(-\sigma)}{p(\sigma)} = e^{\frac{-\tau \sigma}{k}} \qquad (6)$$

Previous derivations of the fluctuation theorem [22, 29] have applied Liouville's theorem to the phase space of open sub-systems in contact with thermal reservoirs, in which entropy production is defined in terms of the rate of phase space contraction of the open sub-system. By applying Liouville's theorem to the combined open sub-system + environment, we obtain the fluctuation theorem in a form consistent with the earlier conjecture of Jaynes [28] based directly on the phase volume conservation of isolated systems. Thus both the 2$^{nd}$ law and the fluctuation theorem can be understood intuitively from figure 1.

## 4. Maximum entropy production from Liouville's theorem

We now resolve the isolated system into an open system plus its environment. Where we draw the boundary between the two is an outstanding issue in applications of MaxEP, see e.g. [21]. The environment is usually chosen to be that part of the entire system detailed knowledge of which is assumed to be irrelevant to the macroscopic state of the open system. For example, this is essentially what we are assuming when we constrain the open system using environmental conditions specified on the boundary only.

Let $C$ denote the stationarity condition and steady-state forcing constraints on the open system, and let $M_C$ denote the set of all macrostates of the open system + environment that are compatible with $C$. For each A ∈ $M_C$ we will write A = $X(A) \oplus E(A)$ where $X(A)$ and $E(A)$ denote the corresponding macrostates of the open system and its environment, respectively. We will denote the corresponding entropy production rate by σ(A), and restrict $M_C$ to those A for which σ(A) > 0 (2$^{nd}$ law), i.e. we are only considering experimentally reproducible macroscopic transitions.

Referring to figure 1, to each possible macrostate A ∈ $M_C$ at $t = 0$ there corresponds a macrostate B = $X(B) \oplus E(B) \in M_C$ into which it evolves over the time interval τ under the imposed constraints C. From the assumption that the open system is in a stationary state, we have $X(B) = X(A) \equiv X$. The corresponding entropy production rate is given by $\sigma(A) = (k/\tau)\ln[W(B)/W(A)]$ and from stationarity we have $\sigma(B) = \sigma(A) \equiv \sigma_X$. Because all microstates in B are macroscopically equivalent, stationarity implies that every microstate in B realizes an open system with stationary macrostate X and entropy production rate $\sigma_X$, i.e. this is true not only for those microstates in the sub-region TA of B that are reachable from A but also, for example, for those microstates in TA′ that are reachable from A′ (figure 1).

But stationarity implies that what is true for B is also true for the pre-image $T^{-1}B$ of B, i.e. for the entire ensemble of microstates at $t = 0$ which end up in B at $t = \tau$. Therefore every microstate in $T^{-1}B$ also realizes a macrostate (A and A′ being two examples) with stationary open system macrostate X and entropy production rate $\sigma_X$. Because each of these macrostates at $t = 0$ has the same entropy production rate $\sigma_X = (k/\tau)\ln[W(B)/W(A)]$, and each of them ends up at $t = \tau$ in the same macrostate B with phase volume $W(B)$, they must each have the same phase volume $W(A)$ as A. Thus, for example, in figure 1 stationarity implies $W(A′) = W(A)$.

It follows that $T^{-1}B$, which by Liouville's theorem has phase volume $W(B)$, is the union of $N_X$ distinct equal-volume macrostates, where

$$N_X = \frac{W(B)}{W(A)} = e^{\frac{\tau \sigma_X}{k}} \qquad (7)$$

Because each of these $N_X$ macrostates (A and A′ being two examples) is characterised by the same open system macrostate X and entropy production rate $\sigma_X$, they differ only in terms of their environmental macrostates E. In other words, there are $N_X$ distinct environmental histories within which a stationary open system macrostate X with entropy production rate $\sigma_X$ can exist over the interval τ. We might then call $N_X$ the *environmental weight* of open system macrostate X over the interval τ (and $k\ln N_X$ the corresponding *environmental entropy*).

Equation (7) shows that open system stationary states with larger entropy production rates have larger environmental weights. The reason is geometrically obvious from figure 1: the larger the entropy production rate of macroscopic history A → B, the greater the phase volume ratio $W(B)/W(A)$, and hence the greater the number of other macroscopic histories A′ → B that share the same entropy production rate. Liouville's theorem is central to this result.

In the absence of any detailed knowledge of E – and, as remarked above, this is typically what determines where we draw the boundary between the open system and its environment – all environmental macrostates consistent with the constraints C are equally probable. The implication of equation (7), then, is that the most probable open system stationary state $X^*$ is the one with the largest entropy production rate, because this $X^*$ can exist within the greatest number of environmental histories. If X and Y are two open system stationary states consistent with the constraints C, the ratio of their environmental weights is

$$\frac{N_X}{N_Y} = e^{\frac{\tau(\sigma_X - \sigma_Y)}{k}} \qquad (8)$$

showing that, as τ → ∞, the MaxEP state $X^*$ becomes the most probable stationary state of the open system by an overwhelming margin.

## 5. Implications for a more general derivation of MaxEP from MaxEnt

The above derivation of MaxEP from Liouville's theorem is heuristic, in the sense that we have assumed that phase space can be divided into macrostates within each of which the corresponding microstates are equally probable. Statistical fluctuations about the 2$^{nd}$ law have also been ignored. In this final section we note how the results of sections 3 and 4 suggest a possible approach to a more fundamental understanding of MaxEP based on MaxEnt.

As Jaynes noted [26], there is an asymptotic link between the Gibbs-Shannon entropy $H = -\sum_i p_i \ln p_i$ and phase volumes $W$, called the Asymptotic Equipartition Property (AEP) of information theory [30]. Specifically, for a given distribution $p_i$, as the number of degrees of freedom $n$ tends to infinity, $H/n$ and $\log W_{HPR}(\varepsilon)/n$ tend to the same limit, where $W_{HPR}(\varepsilon)$ is the phase volume of a "high-probability region" of phase space, defined such that $1-\varepsilon$ is the probability that the system microstate lies in the HPR (surprisingly, the asymptotic result is independent of $\varepsilon$). The AEP generalises Boltzmann's formula $S = k\log W$ to the case where $p_i$ varies continuously over phase space, as it does in MaxEnt, instead of $p_i$ being constant within $W$. The result provides an intuitive interpretation of MaxEnt which amounts to finding the largest possible HPR consistent with the available information.

This asymptotic link between $H$ and phase volume (the AEP or 'generalised Boltzmann formula'), together with the phase volume explanation of MaxEP presented here, suggests there exists a more fundamental link between MaxEnt and MaxEP. The heuristic proofs of the 2$^{nd}$ law and MaxEP based on phase volumes might then be generalised to reveal a more fundamental understanding of these principles based on MaxEnt.

A clue as to how such a derivation of MaxEP from MaxEnt might proceed is given by equations (4) and (7), which together imply the correspondence

$$N_X = \frac{p_\Gamma}{p_{\Gamma*}} \tag{9}$$

This correspondence tentatively suggests that maximising $k\ln N_X$ (the environmental entropy) in the phase volume picture might be equivalent to maximising the expectation value $\langle \ln(p_\Gamma / p_{\Gamma*}) \rangle$ over $p_\Gamma$ in MaxEnt. Therefore, a MaxEnt-based derivation of extremal properties of $\langle \ln(p_\Gamma / p_{\Gamma*}) \rangle$ may lead to a deeper understanding of MaxEP. MaxEnt also opens the way to interpreting the phase volume entropy production rate $\langle \sigma_\Gamma \rangle$ physically in terms of thermodynamic entropy production or dissipation [10]. This approach is currently being studied and will be reported on elsewhere [31].

### Acknowledgements

Financial support from the Isaac Newton Institute for Mathematical Sciences (Programme CLP: *Mathematical and Statistical Approaches to Climate Modelling and Prediction*, Aug-Dec 2010; to RD) and Fondazione Cariparo (to AM) is gratefully acknowledged.